\documentclass[aps,prl,twocolumn,groupedaddress]{revtex4}

\begin{document}
\tighten
\bibliographystyle{prsty}
\title{Do Large-Scale Inhomogeneities Explain Away Dark Energy? }
\author{Ghazal Geshnizjani}\email{ghazal@physics.wisc.edu} \affiliation{Department of Physics, University of Wisconsin,
  Madison, WI 53706}
\author{Daniel J.H. Chung}\email{Danielchung@wisc.edu} \affiliation{Department of Physics, University of Wisconsin,
  Madison, WI 53706} \author{Niayesh
Afshordi}\email{nafshordi@cfa.harvard.edu}\affiliation{Institute
for Theory and Computation, Harvard-Smithsonian Center for
Astrophysics, MS-51, 60 Garden Street, Cambridge, MA 02138}
\date{\today}
\preprint{hep-th/0503553}

\begin{abstract}
Recently, new arguments \cite{Riotto,Kolb:2005me} for how
corrections from super-Hubble modes can explain the present-day
acceleration of the universe have appeared in the literature.
However, in this letter, we argue that, to second order in spatial
gradients, these corrections only amount to a renormalization of
local spatial curvature, and thus cannot account for the negative
deceleration. Moreover, cosmological observations already put
severe bounds on such corrections, at the level of a few percent,
while in the context of inflationary models, these corrections are
typically limited to $\sim 10^{-5}$. Currently there is no general
constraint on the possible correction from higher order gradient
terms, but we argue that such corrections are even more
constrained in the context of inflationary models.
\end{abstract}
\maketitle

\section{Introduction}
The potential impact of large scale perturbations on local
cosmology has been a subject of interest in different contexts
during the past decade \footnote{For some literature discussing
this subject see
\cite{MAB,ABM,WT,Nambu2,aw0,unruh,GB1,niayesh,Geshnizjani:2003cn,Geshnizjani:2004tf}
and references there in.}. In the standard theory of cosmological
perturbations, it is always assumed that perturbations do not have
any impact on the evolution of background cosmology.  However,
there is no a priori reason for neglecting the back reaction from
perturbations, particularly since Einstein's equations are highly
nonlinear. Although it is easy to see that corrections do exist,
an important question is whether these corrections, which are of
second order or higher in perturbations, will ever become
significant and if they do, what is the right way to distinguish
the real physical effects from the gauge ambiguities in the
calculations.
The key difficulty is that if the perturbation effects are
misinterpreted, one may overlook physical bounds already existing
on such effects because the physical bounds are written in terms
of variables not manifestly connected with the perturbations.

Recently \cite{Riotto,Kolb:2005me} argue that corrections due to
the interplay between IR modes and UV modes may lead to an
apparent late time acceleration of the universe, with no need for
dark energy or a cosmological constant.  One of the correction
terms that is claimed to determine the apparent acceleration is of
the form $\varphi\nabla^2 \varphi$, where $\varphi$ is the
gravitational potential. It is argued that this correction can
have a large variance and its statistical nature may cause a
negative value for the observed deceleration parameter.

Even without computation, one might guess that there is a problem
with this correction becoming significant from a phenomenological
point of view. For scale-invariant fluctuations of $\varphi$, the
variance in $\varphi\nabla^2 \varphi$ scales as $\lambda^{-2}$,
where $\lambda$ is the physical length scale.  Therefore, if this
correction is indeed $\sim 1$ on present-day Hubble scale to
explain away dark energy, it will be $ \gg 1$ on smaller scales,
which undermines the incredible success of linear structure
formation theory in the low-redshift universe (see e.g.,
\cite{seljak}).

In this paper, we investigate a related problem.  In the following
sections we will demonstrate that the perturbative corrections of
the form $\varphi\nabla^2 \varphi$ cannot lead to a negative
deceleration parameter (at least not in the manner suggested in
\cite{Riotto,Kolb:2005me}), because this effect stems from a
renormalization of the local spatial curvature \cite{lam}. We then
argue how current cosmological observations put severe constraints
on the magnitude of these corrections. Finally, we discuss the
loopholes (e.g.~our neglect of higher than second order gradients)
in our argument before concluding.

\section{Corrections to deceleration parameter due to spatial curvature}

The metric of a homogeneous and isotropic universe
(Friedmann-Robertson-Walker; FRW metric) can be generally
described as
\begin{equation} \label{Hmetric} ds^2=-dt^2+a^2(t)(1+{1\over4} {\cal
K}r^2)^{-2}\delta_{ij}dx^idx^j,
\end{equation}
where $a$ is the scale factor and ${\cal K}$ is the spatial
curvature. It is customary to normalize ${\cal K}$ such that it
takes the values of $0$, $+1$, and $-1$ corresponding respectively
to flat, closed, and open universes \footnote{Note that, despite
the common nomenclature, the correspondence between the local
curvature and topology of the spatial hypersurface is anything but
unique.}, but in general it could take any value. Einstein's
equations for the above metric reduce to the Friedmann equations
\begin{eqnarray}
\label{Friedmanneqs} H^2+{{\cal K}\over a^2}&=&~~{8\pi G \over 3}~ \rho, \\
{\ddot{a}\over a}&=&-{4\pi G\over 3}(\rho+3p), \label{eq:fried2}
\end{eqnarray} where dot denotes time derivative and $H=\dot{a}/a$, is
the Hubble constant, while $\rho$ and $p$ are respectively the
total energy density and pressure of matter components in the
universe. The deceleration parameter $q$ which describes the
deceleration of the scale factor $a(t)$ is defined as
\begin{equation} q=-{\ddot{a}\over aH^2}. \end{equation} In a
matter dominated universe, where $p=0$, Eqs.~(\ref{Friedmanneqs})
and (\ref{eq:fried2}) imply that
\begin{equation} \label{q(k)}
q={1\over 2}\bigl(1+{{\cal K}\over a^2 H^2}\bigr ).
\end{equation}

\section{Renormalization of the local curvature due to large-scale
inhomogeneities}

In this section we calculate the corrections to the local spatial
curvature, ${\cal K}$, in a flat universe due to large scale
inhomogeneities. To obtain these corrections, we will start by
expanding the metric to second order in perturbations in the
synchronous gauge, the same metric that Barausse et. al. use in
\cite{Riotto}, and then express all perturbative corrections in
terms of the peculiar gravitational potential $\varphi$, as they
did, while dropping all the terms of order higher than $\nabla^2
\varphi$ in the gradient expansion (these are subdominant in the
IR -- i.e. long wavelength-- limit):
\begin{eqnarray} \label{pmetric}
ds^2&=&-a^2d\eta^2+a^2\gamma_{ij}dx^idx^j,\\
\gamma_{ij}&=&(1-{10\over
3}\varphi-{\eta^2\over9}\nabla^2\varphi+{50\over9}\varphi^2+{5\eta^2\over54}\varphi^{,k}\varphi_{,k})\delta_{ij}
\nonumber \\ &-&{\eta^2\over
3}\bigl(\varphi_{,ij}-{1\over3}\delta_{ij}\nabla^2\varphi\bigr)\nonumber
\\&-&{5\eta^2\over9}\bigl(\varphi^{,i}\varphi_{,j}-{1\over3}\varphi^{,k}\varphi_{,k}\delta_{ij}\bigr).
\end{eqnarray}
We can also write the Taylor expansion of $\varphi$ around its
value at the location of a particular observer. Assuming that
$\varphi$ is isotropic around this location \footnote{The first
order term in $r$, as well as the second-order anisotropic terms
will, in the end, drop out to the lowest order, if one is looking
at physical observables such as luminosity distance which are
averaged over the sky.}, we have
\begin{equation}
\varphi \simeq \varphi_0+{1\over 6}\nabla^2 \varphi ~r^2.
\label{eq:expansion}
\end{equation}
Note that this constant $\varphi_0$ corresponds to the
superhorizon modes of the potential fluctuation.  As we will see,
it is the interaction of these superhorizon modes with $\nabla^2
\varphi$ which leads to a modification of the deceleration.
Eq.~(\ref{eq:expansion}) further simplifies the $\gamma_{ij}$ in
the metric of Eq.~(\ref{pmetric}) into
\begin{eqnarray}\label{gamma}
\gamma_{ij}&=&\bigl\{1-{10\over
3}\varphi_0+{50\over9}\varphi^2_0-{\eta^2\over9}\nabla^2\varphi\nonumber\\&-&{5\over
9}\nabla^2\varphi~r^2+{50\over27}\varphi_0\nabla^2\varphi~r^2\bigr\}\delta_{ij}
.
\end{eqnarray}

We now note that we can renormalize the scale factor $a(\eta)$ to
reproduce the metric in Eq.~(\ref{Hmetric}). This can be done by
taking
\begin{equation}\label{newa}
\tilde{a}(\eta)=a(\eta)\bigl[1-{10\over
3}\varphi_0+{50\over9}\varphi^2_0-{\eta^2\over9}\nabla^2\varphi\bigr]^{1/2},
\end{equation}
leading to the following form for the metric
\begin{equation}\label{fmetric}
ds^2=-dt^2+\tilde{a}^2(\eta)(1-{5\over 9}\nabla^2\varphi~
r^2)\delta_{ij}dx^idx^j,
\end{equation}
where the $\varphi_0\nabla^2\varphi$ terms cancel out and we have
ignored the higher order terms. For small values of the curvature
${\cal K}$, the above metric is equivalent to the metric of
Eq.(\ref{Hmetric}), where the curvature term is now
\begin{equation}\label{K}
{\cal K}=
{10\over9}\nabla^2\varphi+O\left[\varphi_0^2\nabla^2\varphi,(\nabla^2\varphi)^2\right].
\end{equation}
\section{Impact of large scale inhomogeneities on the deceleration parameter}
We now compute the correction to the deceleration parameter $q$
due to the renormalization of the local spatial curvature ${\cal
K}$. Substituting from Eq.(\ref{K}) into Eq.(\ref{q(k)}), we find
\begin{equation}\label{q(varphi)}
q={1\over2}\bigl(1+{10\over9}{\nabla^2 \varphi \over
\dot{\tilde{a}}^2}\bigr). \label{qdelphi}\end{equation} Using
Eq.(\ref{newa}), we find
\begin{equation}
\dot{\tilde{a}}^2=\dot{a}^2\bigl[1-{10\over
3}\varphi_0-{\eta^2\over9}\nabla^2\varphi
-\frac{2}{9}\frac{a}{\dot{a}} \eta \nabla^2 \varphi
+\mathcal{O}(\varphi^2)  \bigr]. \label{adot}\end{equation}
Substituting Eq.(\ref{adot}) into Eq.(\ref{qdelphi}), we end up
with corrections to the deceleration parameter
\begin{eqnarray}
&q=&{1\over2}\bigl[1+\bigl({5\over18}\nabla^2
\varphi+{25\over27}\varphi_0\nabla^2\varphi\bigr)\bigl({2\over\dot{a}}\bigr)^2\bigr]\nonumber\\&+&
O\left[\varphi_0^2\nabla^2\varphi,(\nabla^2\varphi)^2\right]
\end{eqnarray}
(neglecting derivatives larger than second order).

Notice that the corrections to $q$ (in particular the third term
including the coefficient), are the exact same correction that
\cite{Riotto} and \cite{Kolb:2005me} argue have statistical nature
and could possibly be the reason for the apparent current
acceleration of the universe.  {\bf However, our result implies
that this correction arises due to the renormalization of the
local spatial curvature, which in nature can never lead to an
acceleration of universe}  (Note that as long as energy density is
positive semidefinite, $1+\mathcal{K}/(a H)^2 \geq 0$).
Furthermore, WMAP \cite{Spergel:2003cb} constraints on
$\Omega_{\cal K}$ (based on the location of CMB Doppler peaks)
lead to a bound on the magnitude of these corrections:
\begin{equation}\label{consWMAP}
\triangle q={1\over2}\Omega_{\cal K}={{\cal K} \over
2\dot{\tilde{a}}^2}< 0.02.
\end{equation}

An even more severe constraint on the magnitude of these
corrections is obtained in the context of inflationary models,
which predict near scale-invariant power spectra of
inhomogeneities. We notice that matter overdensity in a flat
universe on large scales is in fact equal to $\Omega_{\cal K}$.
Therefore, we find
\begin{equation}
\langle \Delta q ^2 \rangle = \frac{1}{4} \Omega_{\cal K}^2 =
\frac{1}{4} \Delta^2_m \simeq 10^{-10} {\rm ~~ on~Hubble~scale},
\end{equation}
where the amplitude of matter overdensities for a scale-invariant
power spectrum, $\Delta_m \sim 10^{-5}$, is observed in a host of
cosmological observations (see e.g., \cite{Spergel:2003cb}).

\section{Discussion}

One deficiency of our argument is that we neglect higher order
spatial gradients \footnote{We thank Toni Riotto for bringing this
issue into our attention.}. Unlike the second order gradients,
which in the long wavelength limit correspond to the local spatial
curvature, there is no property of the homogeneous universe which
can be (without averaging) associated with higher order spatial
gradients of the metric perturbations.  Hence, in principle, there
may be a way to nonperturbatively arrange large averaged
correction to $\rho+3 p$ without disturbing $\mathcal{K}$
significantly\footnote{It has been previously shown that
perturbative UV back-reaction also leads to a renormalization of
${\cal K}$, but its magnitude is again severely constrained by
observations in the regime that perturbation theory can be trusted
\cite{Futamase2}. For more discussion regarding UV back-reaction
see \cite{Futamase1,Futamase3,Ruus,Jelle,Buchert,BE,TF}}.

In other words, if we integrate out UV degrees of freedom except
modes with wave vector of order $H_0$, the renormalization to
$\rho+3 p$ may be significant without significantly perturbing
$\mathcal{K}$ \cite{Rasanen}. Note that one need not integrate out
IR degrees of freedom because approximate homogeneity and isotropy
on cosmological scales of our Hubble patch is consistent with all
observations.  Of course, if the universe is extremely
inhomogeneous outside of our horizon, we must also integrate out
IR modes to reduce the approximate degree of freedom to $\rho$,
$p$, and $\mathcal{K}$ \footnote{Integrating out IR modes is
qualitatively different from integrating out UV modes since
nonlocal terms generated by IR modes do not decouple as the
expansion of the universe causes the theory to flow to IR.}.
Furthermore, a dynamical IR cutoff always exists due to the
existence of a Hubble horizon. Despite this caveat, one
unequivocal point of this paper is that this effect of
renormalizing $\rho+3p$ without disturbing $\mathcal{K}$ must
occur through a pathologically nonuniformly convergent series or
nonperturbative behavior (e.g., without resorting to derivative or
small potential expansion) since perturbatively, the second
gradient order term contributes to the spatial curvature which by
itself cannot account for the acceleration of the universe (and is
severely constrained observationally).

To reemphasize the need for nonperturbative corrections to have a
possibility at explaining the acceleration of the universe, we can
estimate the observational bounds on higher gradient order terms
(assuming derivative expansions to be valid) in the context of
inflationary models with near scale-invariant power spectra. For
scale-invariant perturbations in $\varphi$, higher order
corrections in the gradient expansion take the form
\begin{equation}
\Delta_{n} q \sim \varphi \nabla^{2n} \varphi \propto
\lambda^{-2n},
\end{equation}
where $\lambda$ is the physical scale at which $\Delta_n q$ is
observed. However, fluctuations of $q=0.5 \rho/\bar{\rho}_c
\propto \rho/H_0^2$ are well measured at sub-Hubble scales of say
$\sim 50 {\rm~Mpc}$ (the scale of galaxy surveys) \cite{Tegmark}
\begin{equation} \Delta_n q|_{50 {\rm~Mpc}} < \Delta_{tot} q|_{50
{\rm~Mpc}} = \frac{1}{2} \Delta_m|_{50 {\rm~Mpc}} \sim 0.1
\end{equation}
where $ \Delta_m=\delta \rho_m/\rho_m$ is the matter density
fluctuation.  Therefore, on Hubble scales ($\lambda = H^{-1} \sim
5000 {\rm~Mpc}$) we find
\begin{equation}
\Delta_n q|_{H^{-1}} \sim \left(H^{-1} \over
50{\rm~Mpc}\right)^{-2n} \Delta_n q|_{50 {\rm~Mpc}} \lesssim
10^{-5-4(n-1)},
\end{equation}
and thus, at least to this order of approximation, the corrections
due to higher order terms are even more constrained.  Note that
averaging procedure will generically give the same order of
magnitude as long as the process is perturbative.  Hence, the
nonlinear corrections appear to have a chance of explaining the
acceleration of the universe only if nonperturbative (or
pathological) effects take place.


\section{Conclusion}

We computed the corrections to the local spatial curvature due to
large scale perturbations (up to second derivative expansion) and
showed that they are the same corrections that \cite{Riotto} and
\cite{Kolb:2005me} suggest may lead to the acceleration of
universe (as far as second derivative corrections are concerned).
We conclude that as attractive as it may seem to have
inhomogeneities resolve the dark energy problem, unfortunately,
this term is insufficient due to the fact that spatial curvature
can never lead to an acceleration of the universe (with energy
density positive semi-definite). Furthermore, there are already
severe bounds on this correction, implied from various
cosmological observations, which indicate that this not only
cannot serve as an alternative to the dark energy but also cannot
change the value of the observed deceleration parameter
significantly.  Because our arguments are based on expanding the
metric to second perturbative order in inhomogeneities and
restricting to second order in derivative expansion, one way to
evade
these arguments is through nonperturbative effects. \\

\begin{acknowledgments}

NA wishes to thank the Physics department at the University of
Wisconsin-Madison for its hospitality.  We thank Robert
Brandenberger and Toni Riotto for helpful comments.

\end{acknowledgments}


\begin{thebibliography}{9}
\bibitem{Riotto}
E. Barausse, S. Matarrese, and A. Riotto, astro-ph/0501152.
\bibitem{Kolb:2005me}
  E.~W.~Kolb, S.~Matarrese, A.~Notari and A.~Riotto,
  arXiv:hep-th/0503117.
\bibitem{Spergel:2003cb}
  D.~N.~Spergel {\it et al.}  [WMAP Collaboration],
  Astrophys.\ J.\ Suppl.\  {\bf 148}, 175 (2003)
  [arXiv:astro-ph/0302209].


\bibitem{MAB}
V.~F.~Mukhanov, L.~R.~Abramo and R.~H.~Brandenberger,
Phys.\ Rev.\ Lett.\  {\bf 78}, 1624 (1997) [arXiv:gr-qc/9609026].
\bibitem{ABM}
L.~R. Abramo, R.~H. Brandenberger and V.~F. Mukhanov,
Phys.\ Rev.\ D {\bf 56}, 3248 (1997) [arXiv:gr-qc/9704037].
 \bibitem{WT}
N.~C.~Tsamis and R.~P.~Woodard,
Phys.\ Lett.\ B {\bf 301}, 351 (1993);\\
N.~C.~Tsamis and R.~P.~Woodard,
Nucl.\ Phys.\ B {\bf 474}, 235 (1996) [arXiv:hep-ph/9602315].
\bibitem{Nambu2}
Y.~Nambu,
Phys.\ Rev.\ D {\bf 63}, 044013 (2001) [arXiv:gr-qc/0009005].
\bibitem{aw0}
L.~R.~Abramo and R.~P.~Woodard,
Phys.\ Rev.\ D {\bf 60}, 044010 (1999) [arXiv:astro-ph/9811430].
\bibitem{unruh}
W.~Unruh, ``Cosmological long wavelength perturbations,''
arXiv:astro-ph/9802323.
                                \bibitem{GB1}
G.~Geshnizjani and R.~Brandenberger,
Phys.\ Rev.\ D {\bf 66}, 123507 (2002) [arXiv:gr-qc/0204074].
\bibitem{niayesh}
N.~Afshordi and R.~Brandenberger,
Phys.\ Rev.\ D {\bf 63}, 123505 (2001) [arXiv:gr-qc/0011075].

\bibitem{Geshnizjani:2003cn}
  G.~Geshnizjani and R.~Brandenberger,
  arXiv:hep-th/0310265.



\bibitem{Geshnizjani:2004tf}
  G.~Geshnizjani and N.~Afshordi,
  JCAP {\bf 0501}, 011 (2005)
  [arXiv:gr-qc/0405117].

\bibitem{seljak}
  U.~Seljak {\it et al.},
  arXiv:astro-ph/0407372.

\bibitem{lam}
  A similar issue was raised in R.~H.~Brandenberger and C.~S.~Lam,
  arXiv:hep-th/0407048.

\bibitem{Futamase2}
  T.~Futamase,
  Phys.\ Rev.\ D {\bf 53}, 681 (1996).

\bibitem{Futamase1}
T.~ Futamase, Phys.\ Rev.\ Lett. {\bf 61}, 2175 (1988).



\bibitem{Futamase3}
T.~ Futamase, Mon.\ Not.\ R.\ Astron.\ Sot. {\bf 237}, 187 (1989).

\bibitem{Ruus}
H.~Russ, M.~H.~Soffel, M.~Kasai and G.~Borner,
Phys.\ Rev.\ D {\bf 56}, 2044 (1997) [arXiv:astro-ph/9612218].

\bibitem{Jelle}
J.~P.~Boersma,
Phys.\ Rev.\ D {\bf 57}, 798 (1998) [arXiv:gr-qc/9711057].

\bibitem{Buchert}
T.~Buchert,
Gen.\ Rel.\ Grav.\  {\bf 33}, 1381 (2001) [arXiv:gr-qc/0102049].

\bibitem{BE}
T.~Buchert and J.~Ehlers,
Astron.\ Astrophys.\  {\bf 320}, 1 (1997)
[arXiv:astro-ph/9510056].

\bibitem{TF}
M.~Takada and T.~Futamase,
Gen.\ Rel.\ Grav.\  {\bf 31}, 461 (1999) [arXiv:astro-ph/9901079].

\bibitem{Rasanen}
  The idea has been in particular advocated in S.~Rasanen,
  JCAP {\bf 0402}, 003 (2004)
  [arXiv:astro-ph/0311257].

\bibitem{Tegmark}
  M.~Tegmark {\it et al.}  [SDSS Collaboration],
  Astrophys.\ J.\  {\bf 606}, 702 (2004)
  [arXiv:astro-ph/0310725].




\end{thebibliography}
\end{document}